%% file: main.tex
\documentclass[Afour,times,doublespace,sageapa]{sagej} 

\usepackage{etoolbox}
\makeatletter
\@ifundefined{bibhang}{}{
  \let\bibhang\relax
}
\makeatother

\usepackage[natbibapa]{apacite}
\usepackage{moreverb,url}

\usepackage[colorlinks,bookmarksopen,bookmarksnumbered,citecolor=red,urlcolor=red]{hyperref}

\newcommand\BibTeX{{\rmfamily B\kern-.05em \textsc{i\kern-.025em b}\kern-.08em
T\kern-.1667em\lower.7ex\hbox{E}\kern-.125emX}}

\def\volumeyear{2024}

\begin{document}

\runninghead{Yash Vekaria et al.}

\title{Auditing the Compliance and Enforcement of Twitter’s Advertising Policy}

\author{Yash Vekaria\affilnum{1}, Zubair Shafiq\affilnum{1} and Savvas Zannettou\affilnum{2}}

\affiliation{\affilnum{1}University of California -- Davis, USA\\
\affilnum{2}Delft University of Technology, NL}

\corrauth{Yash Vekaria\\
University of California -- Davis, USA\\
One Shields Ave, Davis, CA 95616, USA\\
\textit{This work was partially carried out during the author's visit to Max Planck Institute for Informatics, Saarbrucken, Germany}}

\email{yvekaria@ucdavis.edu}

\input{abstract}


\maketitle

\noindent \textbf{Disclaimer: This paper studies adult content in Twitter ads. Therefore, we warn the readers about uncensored mentions of sexually explicit content in the manuscript.}

\input{introduction}

\input{related-works}

\input{methodology}

\input{results}

\input{conclusion}

\theendnotes

\bibliographystyle{apacite}
\bibliography{mybib.bib}

\newpage
\begin{sm}
    Anonymized data: \url{https://osf.io/sk8r2/?view_only=17bb5d6cc75343d38c2ece7f24c1ae90} \\
    
    \noindent \textbf{Language distribution of our dataset}
    \label{appendix:language-distribution}
    
    Our dataset comprises of ads in 57 different languages across the globe. Figure~\ref{fig:language_distribution} depicts the distribution of the top most used languages in the textual content of the ads retained after the 14-day rehydration period. We also compare this distribution with the prevalence of the same languages in the random 1\% data and Twitter ads before the rehydration. We observe a strongly co-related language distribution maintained in the 1\% data and the ads data for some top languages, namely, Japanese, Spanish, and Turkish.
    
    \begin{figure}[h]
    \centerline{\includegraphics[width=0.9\columnwidth]{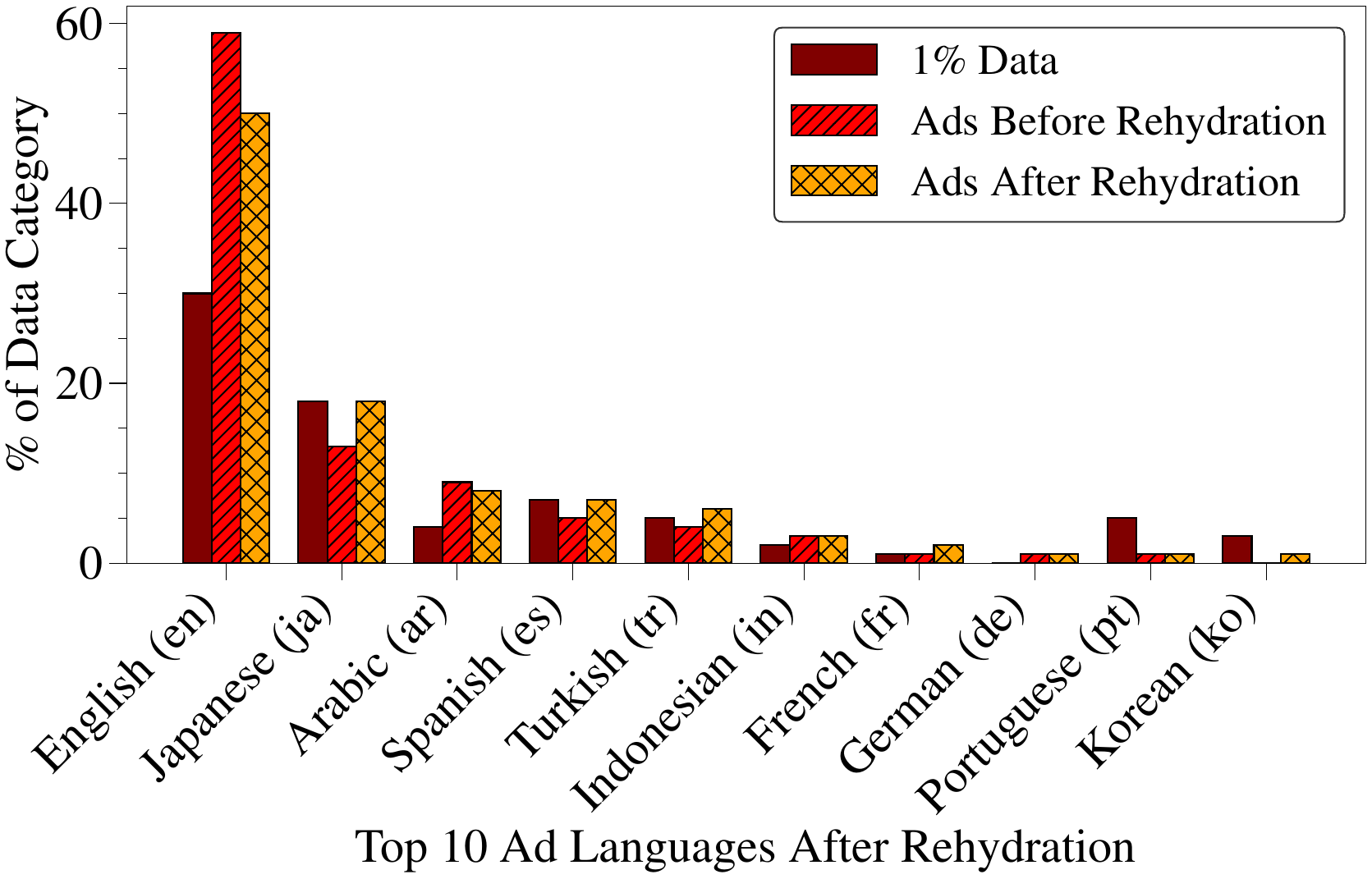}}
    \caption{Distribution of languages in 1\% data as well as in Twitter ads obtained from 1\% data.}
    \label{fig:language_distribution}
    \end{figure}

    \noindent \textbf{Weekly advertising trend of Twitter advertisers}
    
    We try to understand if advertising on Twitter follows a diurnal trend (i.e., advertising is more prevalent on some specific days of the week than others).
    This could be because advertisers, based on their past experiences, have observed receiving higher engagements on their ads if they were shown on some specific day of the week. 
    To look into this diurnal pattern, we plot the average number of total ads and the average number of distinct advertisers advertising on each weekday in Figure~\ref{fig:diurnal_weekdays}. 
    We observe that Twitter advertising decreases over the weekends and then kejpg increasing from Monday onwards until its peak on Friday.
    
    \begin{figure}[h]
    \centerline{\includegraphics[width=0.8\linewidth]{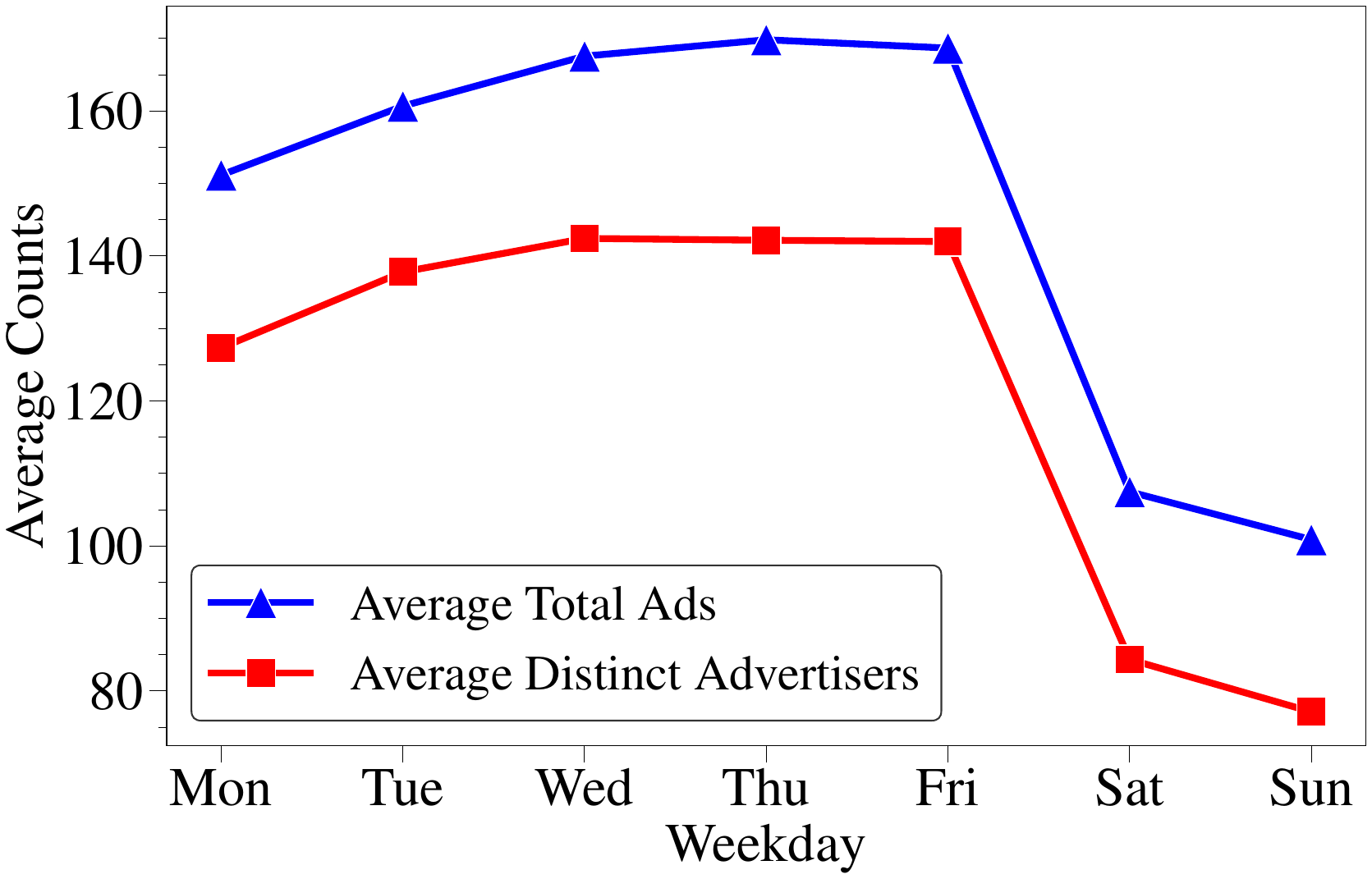}}
    \caption{Diurnal trend of Twitter advertising over week.}
    \label{fig:diurnal_weekdays}
    \end{figure}

    \noindent \textbf{Temporal distribution of ad policy violation}
    
    Now, we show the temporal distribution of violations related adult content advertising policies. Figure~\ref{fig:cdf_all_sex_scores} shows CDF of \texttt{sexually\_explicit} score of perspective API of all refreshed ads in our dataset with positive scores. 
    \begin{figure}
    \centerline{\includegraphics[width=0.78\linewidth]{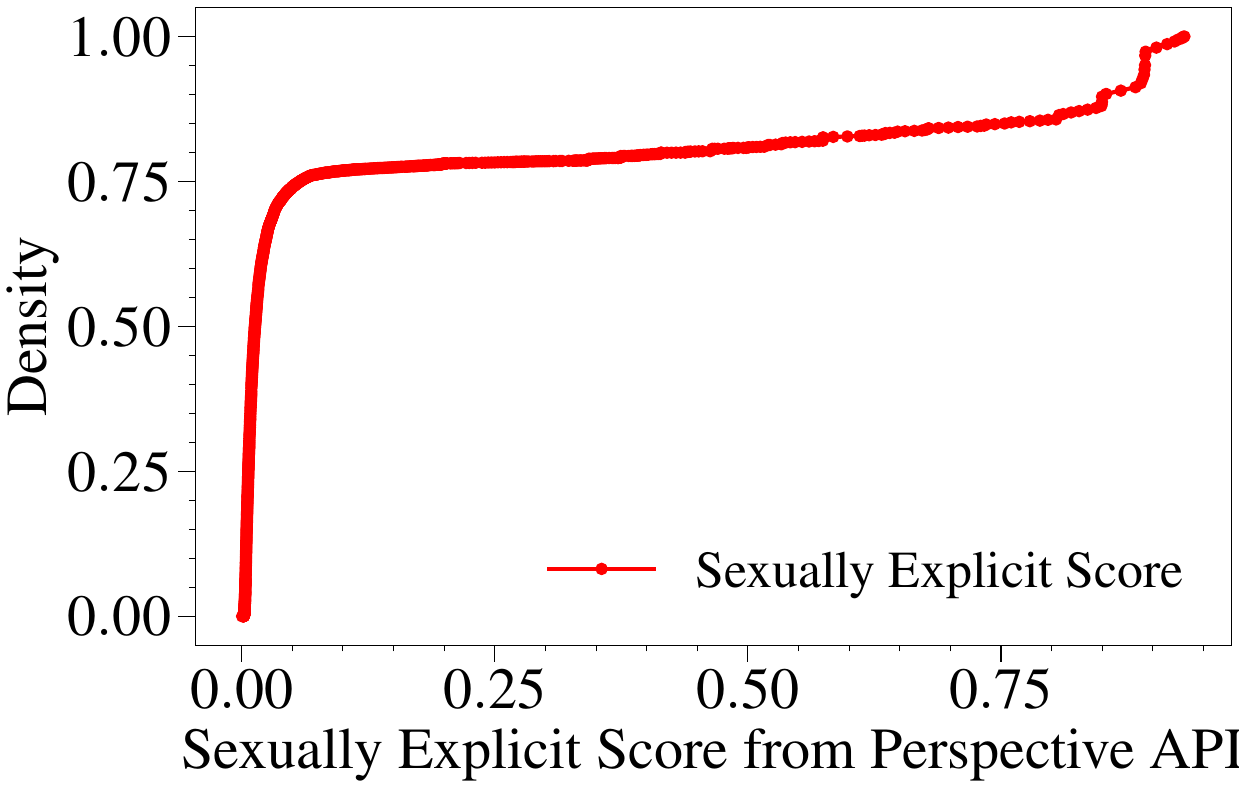}}
    \caption{CDF distribution of the Perspective API scores of 24,530 rehydrated ads in our data for \textit{sexually\_explicit} attribute.}
    \label{fig:cdf_all_sex_scores}
    \end{figure}
    Figure~\ref{fig:sexual_ads_daily_distribution} shows the number of violations for adult ads per day in our dataset. 
    \begin{figure}
    \centerline{\includegraphics[width=0.9\columnwidth]{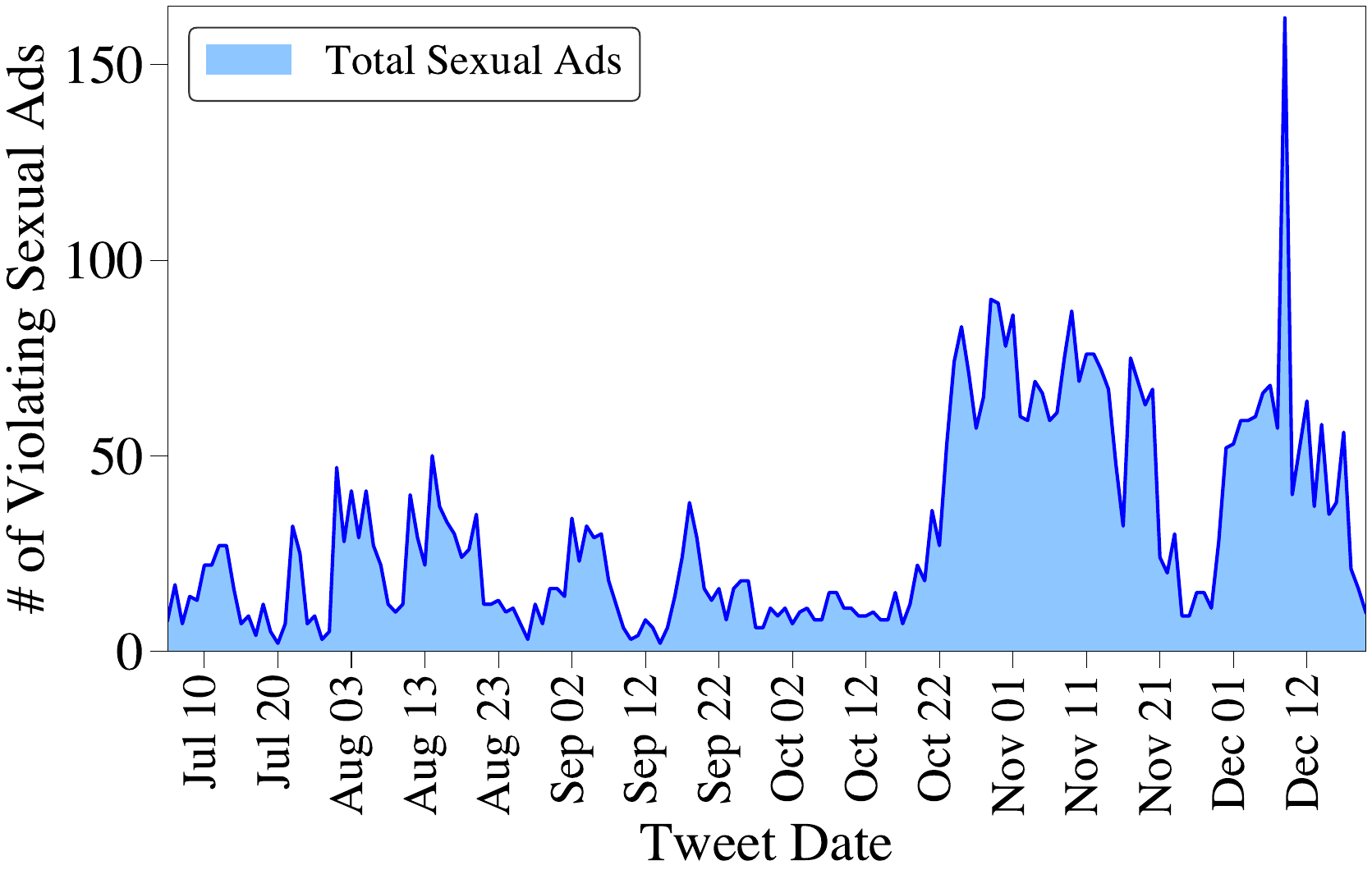}}
    \caption{Daily Distribution of the ads violating Twitter's \textit{adult content ad policy}.}
    \label{fig:sexual_ads_daily_distribution}
    \end{figure}
    We can observe that from late Oct 2023, the number of violating adult ads have significantly increased. 
    One of the plausible reasons could be the layoff of 50\% Twitter employees post  Musk's Twitter acquisition {in multiple rounds}, which included a lay-off of 15\% of moderation staff on Nov 4, 2023~(\cite{twitterlayoffs}). Or it could be simplify due to a general increase Twitter advertising post Musk's takeover as he relaxed platform policies to allow free speech.
\end{sm}

\end{document}

%% file: abstract.tex
\begin{abstract}

Online platforms have enacted various policies to maintain a safe and trustworthy advertising environment. 
However, the extent to which these policies are adhered to and enforced remains a subject of interest and concern. 
In this work, we present a large-scale audit of adult advertising on Twitter (now X), specifically focusing on compliance with its adult (sexual) content advertising policy. 
Twitter is an interesting case study in that it -- uniquely from other social media platforms -- allows posting of adult content but prohibits adult content in advertising.
We analyze approximately 35 thousand ads on Twitter with respect to their compliance to the adult content ad policy through Perspective API and manual annotations.
Among other things, we find that nearly 38\% of ads violate Twitter's adult content advertising policy, although the platform eventually removed only about 63\% of these non-compliant adult ads.
We also find inconsistencies in the moderation of such ads across languages, highlighting the need for more reliable and consistent moderation practices across various languages.
Overall, our findings highlight blind spots in Twitter's adult ad policy enforcement for certain languages and countries. 
Our work underscores the importance of external audits to monitor compliance and improve transparency in online advertising.
\end{abstract}

%% file: introduction.tex
\section{Introduction}
\label{sec: introduction}

Social media platforms allow advertisers to promote their products or services to users on the platform~\citep{FbAdvertising, TwitterAdvertising}.
However, to maintain a safe and trustworthy advertising environment~\citep{geng2021effects}, they impose various policies that all advertisers must adhere to.
For example, social media platforms prohibit harmful advertisements that promote scams, phishing schemes, or illegal products and services~\citep{FbProhibitAds, TwitterProhibitAds}. 
Similarly, social media platforms have strict policies to prevent \textit{sharing} or \textit{promotion} of sexually explicit content via advertising.

Given the potential of online advertising to reach and influence many users~\citep{arora2019measuring}, there is a great deal of interest in understanding whether platforms enforce compliance with their stated advertising policies. 
Some online platforms, such as Facebook, have introduced limited transparency by, for example, making available an archive of the ads that run on the platform~\citep{fbadarchival}. 
However, not all online platforms have committed to such transparency, leading to new regulations around the world~\citep{dsa,dsosa}. 
Even when they do, research has shown that these transparency efforts are far from perfect~\citep{edelson2020security}.
To improve transparency and compliance, there is a pressing need for external audits that provide an independent assessment of advertising policy compliance on online platforms.

In this paper, we conduct the first large-scale audit of compliance and enforcement of adult sexual content advertising policy on Twitter (now X). 
Our time period of analysis consists data from when the platform was Twitter as well as from when it was re-branded as X. 
However, in this paper, we refer to the platform as `Twitter'. 
More than ``another study on Twitter'', through this work, we aim to emphasize the importance of external audits of social media platforms to study their risks and potential online harm.
This is becoming extremely important and relevant as Digital Services Act (DSA) recently went into effect in the EU, mandating large online platforms to aid researchers with appropriate data to assess whether and how their advertising systems are manipulated.
Twitter is an interesting case study since it allows posting of adult content \citep{AdultContentPostingTwitter} but it prohibits adult content in advertising \citep{twitteradultpolicy}. 
Compliance with adult advertising policy is important to avoid exposure of inappropriate content to minors or vulnerable populations~\citep{choi2022sexual,csamstanford2023}.
Anecdotal evidence has also suggested that Twitter is unable to detect and remove sexual exploitation content~\citep{TwitterFailsToRemoveAdult} and that 13\% of Twitter constitutes NSFW content~\citep{13pNSFWTwitter}.
Thus, it is crucial to evaluate the existing gaps and new issues that have emerged in regards to adult advertising policy compliance and enforcement on Twitter more recently than ever before.

Our work aims to answer two research questions:
\begin{itemize}
    \item \textbf{RQ1:} What is the level of moderation for adult sexual advertisements on Twitter?
    \item \textbf{RQ2:} What is the level of compliance for the non-moderated adult advertisements on Twitter with its adult sexual content advertising policy?
\end{itemize}
By `non-moderated' we refer to the ads that are supposedly not detected or removed by Twitter's moderation systems.
To answer these research questions, we collect a dataset of nearly 600 million tweets over six months. 
We then identify tweets that are advertisements by leveraging the tweet metadata made available in Twitter's Streaming API \citep{twitterstreamingapi}. 
We identify nearly 35,000 distinct advertisements by filtering tweets created via Twitter Ad Manager \citep{TwitterAdManager}. 
To identify the tweets that are likely removed by Twitter due to non-compliance, we rehydrate the ads after two weeks. 
We then compare removed and existing ads to understand the effectiveness of Twitter's ad moderation.
To identify advertisements that have a sexually-explicit nature, we use Perspective API \citep{perspectiveapi}. 

\noindent \textbf{Main findings.} Our audit yields the following findings:

\begin{itemize}

\item {We find that 38.64\% of all advertisements on Twitter are sexual in nature and hence violate the adult content advertising policy. 
Twitter moderates 62.68\% of these violating ads; however, 37.32\% of the remaining violating ads exist on the platform even 2 weeks after their creation despite the similarity in their content.}

\item Most of the moderated ads (81\%) are due to a violation of Twitter's adult content advertising policy. 75\% of advertiser accounts whose at least one ad was removed and that were created after the start of our data collection violates Twitter's adult content advertising policy, suggesting spam accounts.

\item Twitter moderates more ads in English, followed by Arabic (24\%) and Indonesian (21.3\%). Also, moderation of Japanese ads (2.8\%) is extremely low as compared to these languages, despite Japanese (31.1\%) being way more frequent than Arabic (8.7\%) or Indonesian (2.6\%).

\item Our analysis of ads that violate Twitter's advertising policies reveals that the majority ($\sim$64\%) of non-moderated adult ads follow a simple ad template comprising a period followed by a set of adult keywords. Advertisers running such ads have a higher tweets-per-minute ratio than a normal account, suggesting automation. Some bad actors are also observed to abuse Twitter ads platform to promote porn content.

\end{itemize}

Overall, ad compliance with policies and enforcement on Twitter remains relatively under-explored, in part because Twitter does not provide an ad library (except recently for European countries~\citep{TwitterAdsTransparency}). 
Our work contributes to the literature by demonstrating how researchers can perform independent and external audits of advertisements on Twitter. 
We anticipate that our proposed data collection methodology can assist researchers in studying the compliance of advertisements with other policies (e.g., political) on Twitter and also motivate similar external audits on other social media platforms.

%% file: related-works.tex
\section{Related Work} 
\label{sec:related-works}

Social media platforms rely on advertising for monetization. 
Given potential negative ramifications, social media platforms are concerned about the abuse of advertising on their platforms. 
Prior research has attempted to understand problematic advertising on the open Web~\citep{zeng2021makes, zeng2021polls, vekaria2022inventory, braun2019fake, kim2019seeing, knoll2016advertising} as well as social media platforms.

\subsection{Ad Transparency on Social Media} \label{related:ad-transparency}
To promote transparency in advertising on their platforms and to comply with the regulations, social media platforms have made efforts to make advertising-related information publicly available. 
In 2019, Facebook started \textit{Ad Library}~\citep{fbadarchival} -- an archive of all the ads that run on their platform. 
Prior work has studied different issues pertaining to advertising on Facebook using this ad archive \citep{chiu2022elaboration, edelson2020security, youn2019newsfeed, abuhashesh2021effect, marino2019emotion}. 
Various studies have explored different types of problematic advertising on Facebook such as discriminatory ad delivery~\citep{ali2019discrimination}, exploitation of sensitive data for ads~\citep{cabanas2018unveiling}, Russia-linked targeting of socially divisive ads~\citep{ribeiro2019microtargeting} and aggravating ads~\citep{vargo2020fear}, COVID-19~\citep{mejova2020covid} or vaccine-related~\citep{jamison2020vaccine} advertising, disparity of harmful ads~\citep{ali2022all}, etc. 

Unlike Facebook, Twitter does not allow direct access to all the ads that run on its platform, hindering at-scale analysis of its advertising ecosystem. 
Twitter also does not have clear guidelines on how to identify ads nor it exposes clear identification suggesting if an ad was paid for or not in their platform data crawled using the API access. 
Due to these limitations, little to no research has focused on studying advertising on Twitter on a large scale. 
Most of the Twitter-specific studies focus on understanding ad-targeting from the marketing perspective on either a limited scale, specific category, or a small set of advertisers~\citep{wei2020twitter, hayes2020advertising, arora2019measuring, mr2022earlyad, wells2020trump, janson2022pandemic, bradley2019major, clark2016vaporous, noh2021relationship}.
Some of these works leverage the tweets available on an advertiser's Twitter timeline as a representative sample for the tweets advertised by them. 
However, this is inaccurate because an advertiser may not always create a tweet for advertising purposes. 
It could also create and share non-ad tweets (i.e., tweets that are not advertised or sponsored), just like a normal Twitter user. 

In contrast to these studies, we propose a novel methodology to effectively detect actual ads from the Twitter metadata as described later in Section~\ref{methodology:ad_identification}. 
As a result, tweets that we detect as ads are truly created from Twitter's advertising platform accessible via an advertiser account, helping us study accurate advertising trends prevalent on Twitter. 
Moreover, our data collection is not advertiser-specific and involves a continuous collection of ads from random 1\% worldwide tweets over a period of 6 months, aiding us to analyze ads on a large scale, as opposed to previous works.

\subsection{Adult Advertising on Social Media}

Past research has studied adult advertising on the Web focused on the promotion of adult services via advertisements by sex traffickers~\citep{keskin2021cracking} and sex workers~\citep{skidmore2018threat}.
Advertising related to adult content is concerning due to the potential exposure of inappropriate content to minors or vulnerable populations. 
For instance, as of 2023, 6.6\% of Twitter's population comprises users in the age range of 13-17~\citep{TwitterUsersbyAge}. Exposure to adult ads due to poor ad moderation can have adverse consequences on children.
Some studies have analyzed normal tweets related to illicit adult content~\citep{burbano2018illicit} and escort services posted by OnlyFans users on Twitter~\citep{arenas2023social} by treating them as advertisements. This clearly shows a lack of understanding in the research community regarding the identification of ads. To this end, we propose and explain the novel methodology in Section~\ref{methodology:ad_identification}.
Thus, no work has really studied adult \textit{advertising} on Twitter, however, research has attempted to detect abusive Twitter accounts~\citep{abozinadah2017statistical, cheng2014isc}, malicious activity~\citep{ilias2021detecting, singh2016behavioral, founta2018large}, and spread of adult content in general on Twitter~\citep{alshehri2018think, mubarak2021adult, dhaka2022detection}. 
To prevent the distribution of such problematic content, extensive research has been conducted in the past to understand content moderation strategies on Twitter~\citep{gorwa2020algorithmic, jimenez2022economics, alizadeh2022content, myers2018censored, papakyriakopoulos2020spread, morrow2022emerging, keller2020facts, jhaver2021evaluating, haimson2021disproportionate, de2022modulating}.
Although Twitter allows sharing normal tweets with adult content, it prohibits the \textit{promotion} of adult content globally. 
This includes:
\vspace{-2mm}
\begin{quote}
    \textit{pornography, escort services and prostitution, full and partial nudity, penis enlargement products \& services and breast enhancement services, modeled clothing that is sexual in nature, dating sites which focus on facilitating sexual encounters or infidelity, dating sites in which money, goods or services are exchanged in return for a date, mail order bride services, sex toys, and host and hostess clubs (Kyabakura)}. 
\end{quote}

\begin{figure*}[t]
\centering
\centerline{\includegraphics[width=0.7\linewidth]{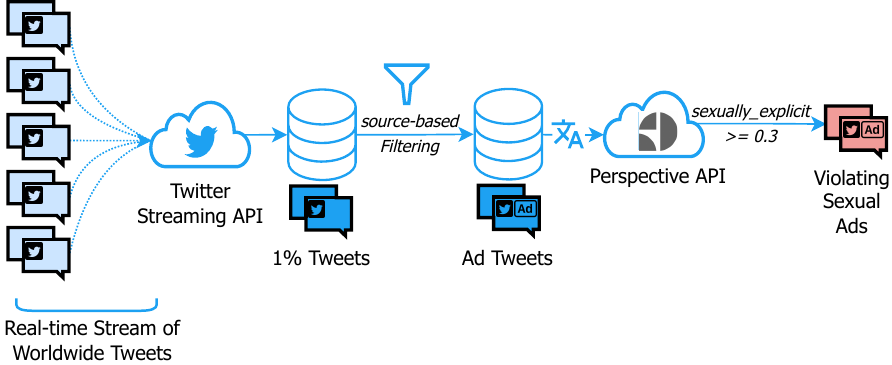}}
\caption{Overview of our methodology.}
\label{fig:our_methodology}
\end{figure*}

This is referred to as adult (sexual) content advertising policy~\citep{twitteradultpolicy}. 
Like Twitter, other social media platforms (including Facebook) also strictly disallows any advertising related to adult content with an exception that allows its inclusion in promotion of art such as paintings or sculptures \citep{fbadultpolicy}.

%% file: methodology.tex
\section{Dataset \& Methodology}
\label{sec:methdology}

This section presents our dataset and research methodology as depicted in Figure~\ref{fig:our_methodology}. 
First, we collect a random 1\% sample of real-time worldwide tweets (Section~\ref{methodology:data_collection}). 
Next, we identify ads observed in the 1\% tweet sample (Section~\ref{methodology:ad_identification}). 
Then, we describe how we identify and study adult ads (Section~\ref{methodology:sexual_ads}). 
Also, we describe how we study potentially malicious URLs embedded in adult ads on Twitter in Section~\ref{methodology:url_analysis}.
Finally, we discuss the ethical considerations when collecting and processing our dataset (Section~\ref{methdology:ethics}).
Data related to our research is made available on the OSF repository (\url{https://osf.io/sk8r2/?view_only=17bb5d6cc75343d38c2ece7f24c1ae90})

\vspace{-0.2cm}
\subsection{Data collection}
\label{methodology:data_collection}

We use Twitter's Streaming API endpoint~\citep{twitterstreamingapi}, which returns a random sample of 1\% of all the tweets made at each given point of time during the day anywhere in the world. 
To gather metadata related to each tweet, we append all available query parameters to the endpoint URL, particularly, \texttt{tweet.fields}, \texttt{user.fields}, \texttt{place.fields}, and \texttt{media.fields}. 
We run our data collection from July 2022 to December 2022, collecting a set of 597,889,636 tweets. 
Then, we identify the tweets that are ads. 
To do this, we leverage the \texttt{source} attribute as discussed in Section~\ref{methodology:ad_identification} and select all tweets with a source in our ad sources, obtaining a set of 34,606 ad tweets. 
Each ad or tweet undergoes Twitter moderation, and if Twitter deems some tweets unsafe, then the account may be temporarily restricted or permanently suspended, making their tweets inaccessible. 
A tweet creator could also delete the tweet if it receives a warning from Twitter. 
To retain ads that are not removed from Twitter, we rehydrate (or refresh) ad tweets after two weeks from their initial fetch date using the Tweet Lookup API~\citep{twitterlookupapi} endpoint: \url{https://api.twitter.com/2/tweets} {as done by Pochat et al.~\citep{le2022audit}}.
Rehydration yielded only 24,530 ads, corresponding to a drop of about 30\% ad tweets.\endnote{We observe that only a small fraction ($<$5\%) of ads are removed by Twitter even after the 2-week rehydration period.}

\subsection{Identifying Ads on Twitter}
\label{methodology:ad_identification}

An inherent challenge that exists when studying online advertising is obtaining data from social media platforms.
Unlike Facebook, Twitter does not provide an ad archive.
To overcome this challenge, in this work, we devise a novel methodology that uses the \texttt{source} attribute on Twitter to identify whether a tweet is an ad or not. 
The \texttt{source} attribute is a metadata field that indicates the application that was used for the posting of the tweet. 
For instance, if a user posts a tweet using Twitter's Web interface, the source will be set to \textit{Twitter Web App}, or if they posted via their iPhone, the source will be set to \textit{Twitter for iPhone}. 
Similarly, the source attribute can be used to identify tweets created via \textit{Twitter's Ads Manager}~\citep{TwitterAdManager}, which assists our purpose of identifying ads.

To advertise on Twitter, an advertiser needs to create a Twitter advertising account, which gives them access to the Twitter Ads Manager to create and manage their ad campaigns. 
A new ad tweet can be created through the advertiser's approved ad account using either the \textit{Twitter's Ads API} (\textit{Twitter Ads} source) or using one of the two options from Twitter Ads Manager page - \textit{Tweet Composer} (under Creatives) and \textit{Create Campaign} (under Campaign). 
All ads created through Tweet Composer are assigned the source \textit{Twitter for Advertisers} or \textit{Twitter for Advertisers (legacy)} label. 
Ads created from the \textit{Create Campaign} option can be either a Simple Campaign (\textit{simpleads-ui} source) or Advanced Campaign (\textit{advertiser-interface} source). 
We use the following sources for identifying ad tweets: \textit{Twitter Ads}, \textit{Twitter for Advertisers}, \textit{Twitter for Advertisers (legacy)}, \textit{simpleads-ui}, and \textit{advertiser-interface}. 
We refer to these as our \textit{ad sources}.

\noindent \textbf{Limitations.}
Our methodology for identifying ad tweets has some limitations that are worth mentioning. 
First, an advertiser can in theory, use a custom source to create and promote their tweets; hence, such tweets will not be captured by our methodology. 
Second, a Twitter user can create tweets via Twitter Ads Manager and post them as regular tweets (unpaid) instead of paid ad tweets. 
Based on our experience with Twitter advertising, we expect this phenomenon to be extremely infrequent. 
Finally, Twitter has recently made changes to its API \citep{TwitterTiersAPIAccess} such that only paid (Pro and Enterprise) versions allow access to the Streaming API. 
Our data collection stopped when we lost access to the Streaming API via our academic account.
That said, our methodology is still applicable -- albeit needing paid API access -- to study advertising on Twitter.

\subsection{Identifying Adult Ads}
\label{methodology:sexual_ads}

To identify adult sexual content in Twitter ads, we use \textit{Google's Perspective API}~\citep{perspectiveapi}. 
Perspective API uses machine learning models to identify different types of abusive comments and the perceived impact of a comment in the conversation. 
It provides scores for \textit{toxicity}, \textit{insult}, \textit{identity} \textit{attack}, \textit{threat}, \textit{profanity}, and \texttt{$sexually\_explicit$} attributes, ranging between 0 to 1 for each attribute. 
This score can be interpreted as the probability of the text being of a specific abusive nature. 
We focused only on \texttt{$sexually\_explicit$} attribute, which captures ``\textit{references to sexual acts, body parts, or other lewd content}'' as broadly scoped by Twitter's definition of adult sexual content.

\begin{figure}
    \centering
    \includegraphics[width=0.95\linewidth]{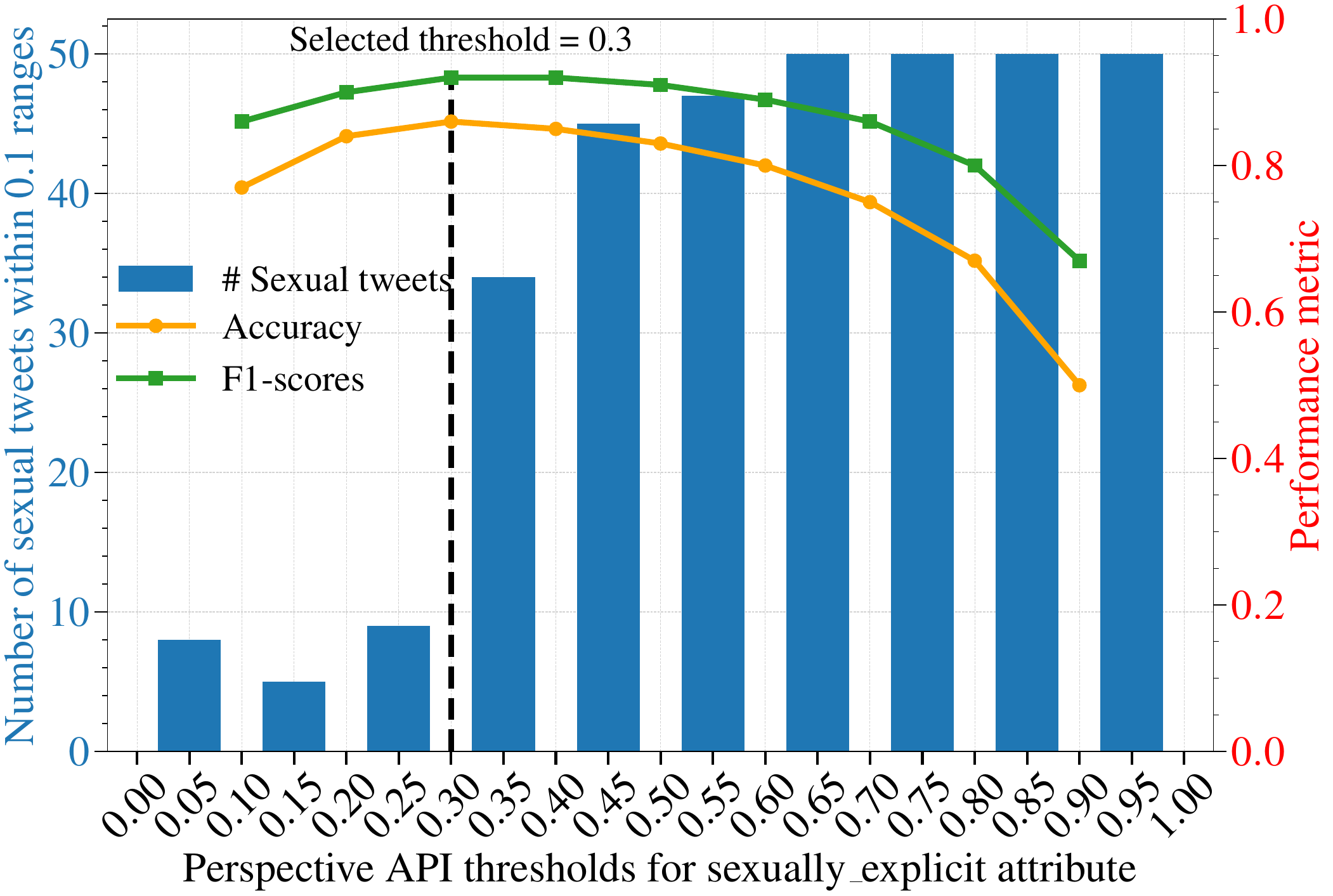}
    \caption{{Threshold selection for Perspective API's \texttt{$sexually\_explicit$} attribute to identify adult ads.}}
    \label{fig:sexually_explicit_threshold_selection}
\end{figure}

The textual content of each ad tweet is first translated to English (if necessary\endnote{$\sim$50\% of the ads are not in English.}) using Google's Translate API~\citep{googletransapi} as popularly used in the literature~\citep{van2020twitter,de2018no,miftari2020analysis} and then passed to the Perspective API to obtain the \texttt{$sexually\_explicit$} attribute score between 0 and 1. 
To decide a suitable threshold for the \texttt{$sexually\_explicit$} attribute, we extract a random sample of 500 ads (50 ad tweets from each 0.1 range of the score from 0 to 1 motivated from~\cite{hoseini2023globalization}). 
A sample of 200/500 ads (20/50 ad tweets from each 0.1 range) is selected to evaluate inter-annotator agreement.
Each ad in this sample is independently annotated to be sexually explicit or not by four researchers (one internal and three external) by referring to the policy documentation. 
We find a 90.5\% inter-annotator agreement and Fleiss' Kappa score of 0.884, which suggests an \textit{almost perfect agreement.}
Next, to evaluate the classification performance, we calculate f1-scores for each 0.1 score range using the full sample of 500 ads. We select the threshold as 0.3 -- for which we observe the highest f1-score (f1-score = 0.92) {and the greatest difference in identified adult ads as depicted in Figure~\ref{fig:sexually_explicit_threshold_selection}}. 
False negatives constitute of ads that contain ``double meaning'' text, which makes the ML model behind the perspective API classify it with a low \texttt{$sexually\_explicit$} score whereas it actually implies sexual references when read by a human.
All the ads with \texttt{$sexually\_explicit$} score $\geq 0.3$ are classified as adult ads. 
We obtain a total of 13,374 \texttt{$sexually\_explicit$} ads from a total of 34,606 ads in our dataset. 
For further validation, we manually verify that these 13,374 ads indeed include or refer to sexually explicit content and are in violation of Twitter's adult sexual content advertising policy; we observe only 314 ($\sim$2.35\%) false positive ads. 
This suggests that only 314 ads are falsely classified as sexually explicit, when they are actually not, showing the effectiveness of using a threshold of 0.3. 
These 314 ads are discarded to obtain 13,060 \texttt{$sexually\_explicit$} ads further studied in Section~\ref{results}. 

\noindent \textbf{Limitations.}
It is impossible to fully account for multiple cultures and customs when analyzing multiple languages. We acknowledge that the annotation is carried out by non-culture experts -- who may not be able to incorporate cultural differences in their annotations.

\begin{figure*}[t]
\begin{center}
\centerline{\includegraphics[width=\textwidth]{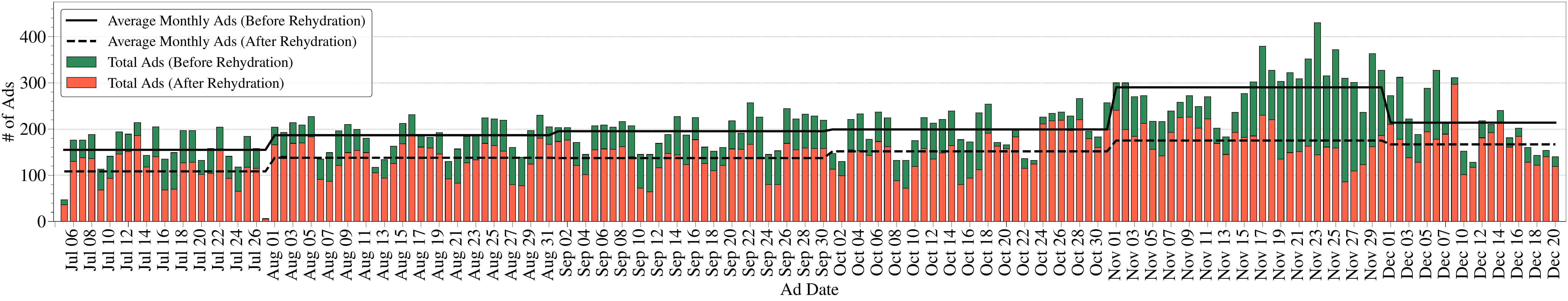}}
\caption{Distribution of ads on Twitter over the period of data collection before and after the rehydration (or refresh).} 
\label{fig:tweet_distribution}
\vspace{-6mm}
\end{center}
\end{figure*}

\subsection{{URL Extraction and Classification}}
\label{methodology:url_analysis}

URLs are often used in ads by advertisers to promote their products or websites. 
Malicious links embedded in the ads can negatively affect many people interacting with them and have largely been observed in tweets with adult content~\citep{10.1145/3442381.3450071}. 
Advances in cybersecurity have automated the detection of such URLs for large-scale applications. 
Hence, it is the responsibility of the platform to also moderate and appropriately handle ads from advertisers that embed and spread malicious URLs. 
While annotating adult ads, we came across several such malicious links. Hence, we study the security of URL usage in Twitter ads in context of the adult ads to understand if Twitter is able to reliably detect and remove ads containing such malicious URLs to enforce their link embedding policy~\citep{twitterlinks}.
We extract embedded links from the tweet metadata obtained from the API response. 
Often, the embedded URLs in the tweet are shortened URLs (e.g., t.co, bit.ly, etc.) and do not actually represent the final landing page that is actually embedded. 
Hence, we visit each of the embedded URL in a fresh browser window programmatically and wait for all the intermediate redirects to complete before extracting the final landing page. 
We check the landing page URL against \textit{VirusTotal}~\citep{virustotal} using their API endpoint to obtain counts for the number of anti-virus services that classify the URL as \textit{malicious} and/or \textit{suspicious}. 
To compare URL usage against non-advertised tweets, we extract a random sample of non-ad tweets that have at least one embedded URL from the 1\% dump of size equal to the total number of ads in our dataset.

\vspace{-2mm}
\subsection{Ethical Considerations}\label{methdology:ethics}
Our work relies solely on analyzing publicly available datasets obtained via Twitter API; hence, we do not deal with sensitive data that is private.
Due to this, no additional privacy concerns arise from our work beyond those that are also applicable to the vast research work that analyzes large-scale datasets obtained via the Twitter API~\citep{jhaver2021evaluating, qayyum2023longitudinal}.
In this work, we performed manual annotations to label advertisements and assess whether they were related to sexually explicit content. 
Out of the four annotators, one was internal (one of the authors), and three were external (two undergrads and one PhD student). 
We acknowledge that this manual work might expose the annotators to sensitive and disturbing content. 
As a result, an informed consent was taken from each annotator by explaining the details of the annotation work and the fact that they might be exposed to such content.
We elected only three external annotators to minimize the exposure to potentially disturbing content to only a few annotators, but also sufficient to establish credibility in our annotations.
On a separate note, when extracting embedded URLs from ads and visiting them independently to extract landing pages, we do not remove campaign details or user identifiers (like `?uid=XX\&etc') (if any). 
Since extraction was done from Tweet metadata with one of the author's API keys, we expect these identifiers to be either absent or associated with the author. 
Moreover, we intentionally do not perform stripping of these identifiers because, in dynamic campaigns, the resultant landing pages could be selected dynamically. These identifiers are essential to accurately replicate the actual behavior of a real user click. 
The crawls were carried out in a containerized environment, and hence, the researchers weren't exposed to any harmful results from clicking malicious URLs.

%% file: results.tex
\vspace{-4mm}
\section{Results}
\label{results}

From our dataset of 34,606 advertisements over a time period of 6 months, 10,306 were removed by Twitter within two weeks of their creation. 
We first discuss the general characteristics of the ads that Twitter is successfully able to detect and remove and contextualize it with respect to adult ad policy violation in Section~\ref{sec:removed-results}. 
We then investigate the compliance of non-moderated ads with Twitter's adult content advertising policy in Section~\ref{sec:rehydrated-results}.

\input{results-removed-ads}

\input{results-rehydrated-ads}

%% file: results-removed-ads.tex
\subsection{RQ1: (In)effectiveness of Twitter's Enforcement}
\label{sec:removed-results}

We first analyze the ad content that Twitter is likely to remove due to violations of its platform policies.
Figure~\ref{fig:tweet_distribution} depicts the distribution of ads per day -- before and after the rehydration during the period of our data collection. 
We observe that 15\%-55\% ($\sim$28\% on an average) of daily ads on Twitter are removed by either the platform or the advertiser. 
The total number of ads in Twitter's 1\% data has gradually increased beginning October 2022 post-Twitter's acquisition, along with an increase in ad removals up to 75\% ($\sim$44\% on an average) daily ads since mid-November 2022.

\begin{figure}
    \centering
    \includegraphics[width=0.96\linewidth]{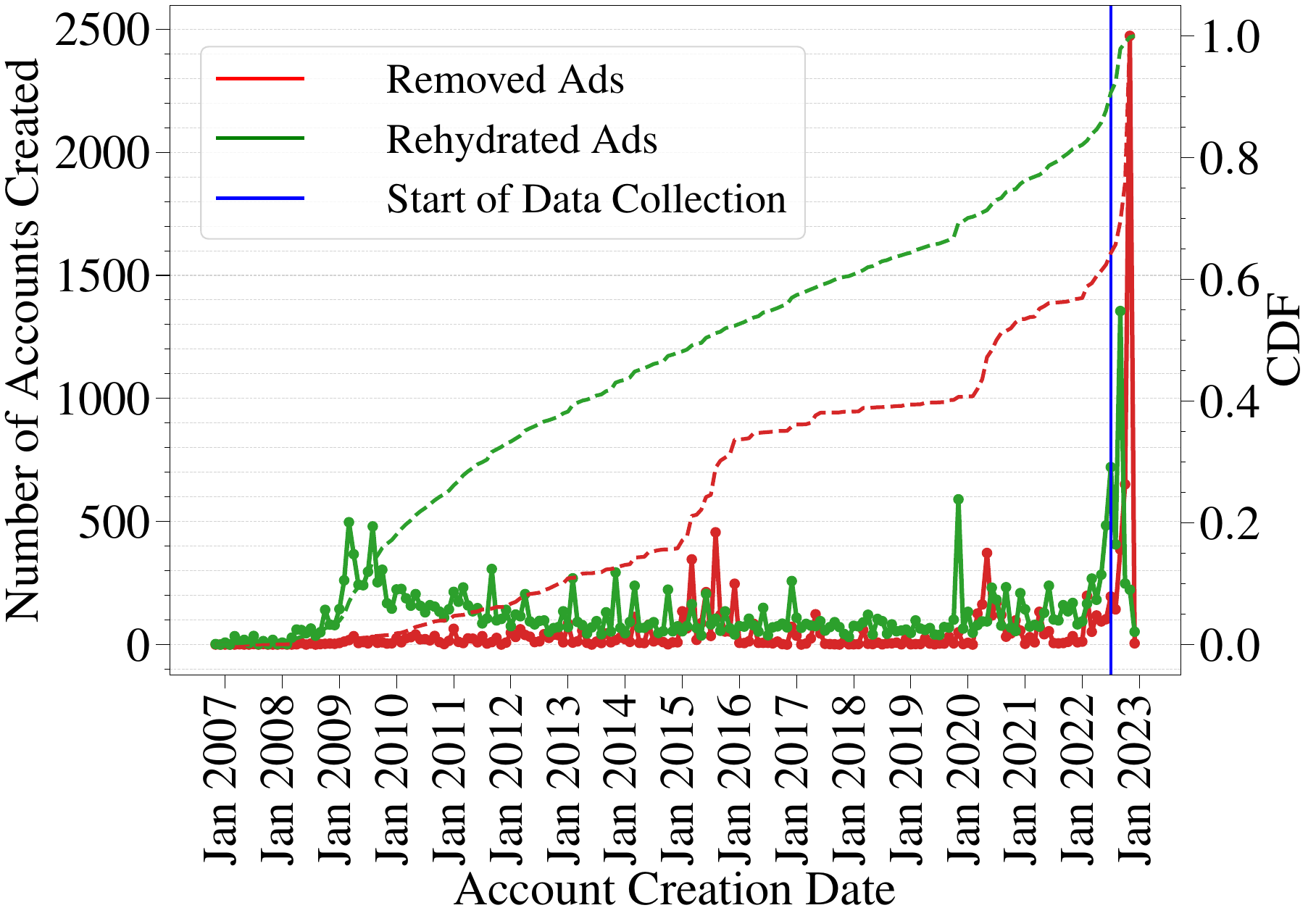}
    \caption{Distribution of account creation dates of advertisers whose ads are removed with respect to not removed by Twitter moderation. Solid lines correspond to number of accounts created while dashed line corresponds to its cumulative distribution function (CDF).}
    \label{fig:removed-ads-account-creation}
    \vspace{-2mm}
\end{figure}

The analysis yields that these ad removals appear to mostly occur from ads campaigned by less popular advertisers --- 50\% of all advertisers with less than 100 followers ($\sim$80\% with less than 10K followers) have at least one advertised tweet removed by Twitter. Some of these advertisers also resemble the characteristics of a spam account with very low follower-to-following ratios. 
One could argue that accounts demonstrating bot-based activity and posting violating ads could have been created after the start of our data collection. To rule out this possibility, we analyze the creation of accounts of advertisers associated with advertising on Twitter. Figure~\ref{fig:removed-ads-account-creation} shows the distribution of account creation dates of all advertisers in our data -- the ones whose at least one ad was removed post-rehydration as well as the ones whose ads were not removed.
Overall, out of 10,221 distinct advertisers in our data, $\sim$80\% accounts were created before the start of our data collection.
In contrast, $\sim$44\% (i.e., 915 out of 2089) advertiser accounts whose ads were removed by Twitter were created after the start of our data collection. 
Moreover, 75\% of 915 advertising accounts promoted at least one ad that violated Twitter's adult content advertising policy. We do not perform analysis of all tweets from each of these accounts as it is out-of-scope for our research. However, this suggests that a majority of ads that are removed from the accounts created after the start of our data collection are due to their violation of Twitter's adult content advertising policy.

Having looked at the general characteristics of ad removals, we analyze the content of the removed ads to identify violations pertaining to adult content as described in Section~\ref{methodology:sexual_ads}.
Out of 10306 removed ads, 8383 ads (i.e., $\sim$81\%) have \textit{sexually\_explicit} attribute score $\geq$0.3 and hence were likely removed due to violation of Twitter's \textit{adult content} advertising policy. 
Surprisingly, these ads also mostly follow a similar content pattern as non-moderated adult ads as described later in Section~\ref{sec:rehydrated-results}. Hence, we perform an in-depth analysis of the content in that section. However, this shows that Twitter is able to detect and remove some adult ads within two weeks of their posting while not others, despite their content being extremely similar.

\begin{figure}[t]
\centerline{\includegraphics[width=0.80\columnwidth]{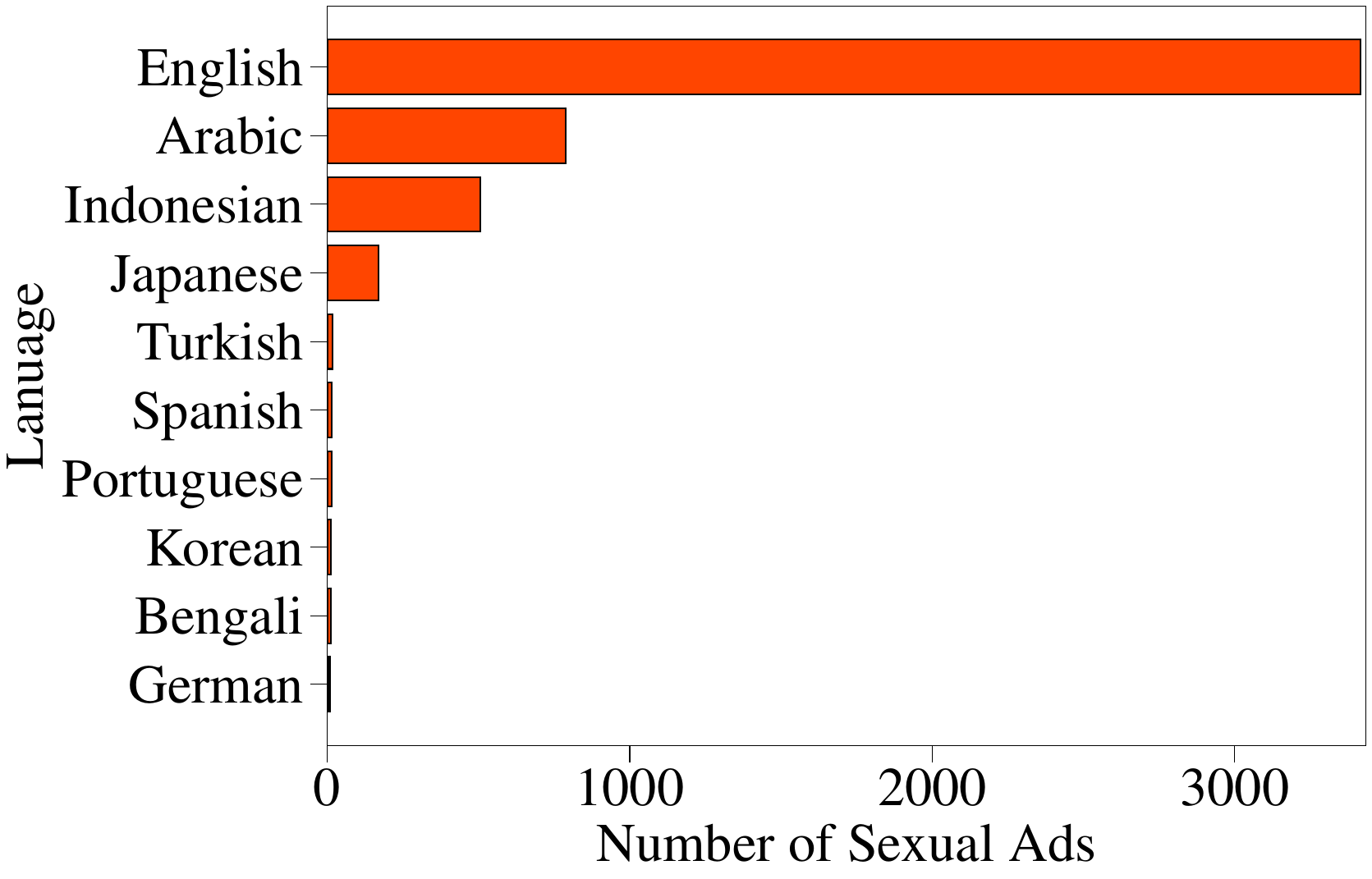}}
\caption{Language distribution of the violating adult ads}
\label{fig:lang_distribution_sex_ads}
\end{figure}

To understand language-specific differences in ad moderation of sexually explicit content, we look at the distribution of language detected in adult ads that are moderated and removed by Twitter.
We observed that out of 8383 ad removals due to violation of adult content advertising policy, cases where an ad contains textual content in English are the highest (8001; 95.4\% of all removed adult ads) followed by Arabic (250; 2.9\% of all removed adult ads), Indonesian (136; 1.6\% of all removed adult ads) and then Japanese (5; 0.06\% of all removed adult ads) (see Figure~\ref{fig:lang_distribution_sex_ads}). The second most prominent language of ads in our entire dataset of 34K ads is Japanese (4533; 13.1\% of all ads), followed by Arabic (3010; 8.7\% of all ads) and Indonesian (908; 2.6\% of all ads) at sixth as observed in Figure~\ref{fig:language_distribution}. However, interestingly, we detect a higher number of ad removals in Arabic and Indonesian as compared to Japanese.
To further corroborate this claim, we compare the distribution of ad removals with policy violations of adult content ad policy in non-moderated ads for these two languages. 
We observe 171 violating Japanese adult ads, 504 violating Indonesian adult ads, and 789 violating Arabic adult ads that are not moderated by Twitter. If we compute the percentage of violating ads that are moderated per language, it turns out to be 2.8\% for Japanese, 21.25\% for Indonesian, and 24\%  for Arabic. This shows that moderation in Japanese is poorer ($\delta$=$21\%$) as compared to Indonesian or Arabic ads, suggesting language-specific differences in Twitter's ad moderation.

\noindent \textbf{Takeaways.}
The key takeaways from our analyses of ads removed by Twitter are as follows:
\begin{itemize}
    \item On average, Twitter observes a removal of $\sim$28\% of daily ads due to policy violations that increase to $\sim$44\% post-Twitter's acquisition.
    \item 75\% of advertiser accounts whose at least one ad was removed and that were created after the start of our data collection violates Twitter's adult content advertising policy, suggesting spam accounts.
    \item $\sim$81\% of the removed ads (i.e., 8383 ads) violate Twitter's adult content ad policy. This suggests that Twitter's ad moderation is effective in detecting policy violations related to the platform's adult content advertising policy.
    \item We find some blindspots of Twitter's enforcement -- Twitter is (in)effective at removing non-compliant ads in specific languages -- we observe Twitter to moderate more ads in English, followed by Arabic and Indonesian. Also, moderation in Japanese is extremely low compared to these languages, even though Japanese is way more frequent than Arabic or Indonesian among all ads.
\end{itemize}

%% file: results-rehydrated-ads.tex
\subsection{RQ2: Compliance of Non-Moderated Ads with Twitter's Adult Advertising Policies}
\label{sec:rehydrated-results}

To better understand the blindspots of Twitter's enforcement, we now analyze the compliance of non-moderated ads with Twitter's adult content advertising policy.  
By `non-moderated,' we refer to these ad tweets that are supposedly not detected or removed by Twitter's moderation systems.
Using the methodology described in Section~\ref{methodology:sexual_ads}, we identify 4,991 ($\sim$20.35\%) ads out of 24,530 rehydrated ads that have a \textit{sexually\_explicit} score of 0.3 or above and manually remove 118 false positive ads to obtain 4,873 \textit{sexually\_explicit} ads. 
We find more than 75\% of ads in our data have \textit{sexually\_explicit} score of 0.05 or less, while $\sim$10\% of ads have scores as high as $>$0.9 (Figure~\ref{fig:cdf_all_sex_scores}).
Out of 34,606 ads in the entire dataset, 13,256 sexually explicit ads were obtained -- 63.24\% of which are already moderated and removed by the platform, as discussed in the previous section. We analyze the remaining 4873 (36.76\%) ads here.

\begin{figure}
    \centering
    \includegraphics[width=\linewidth]{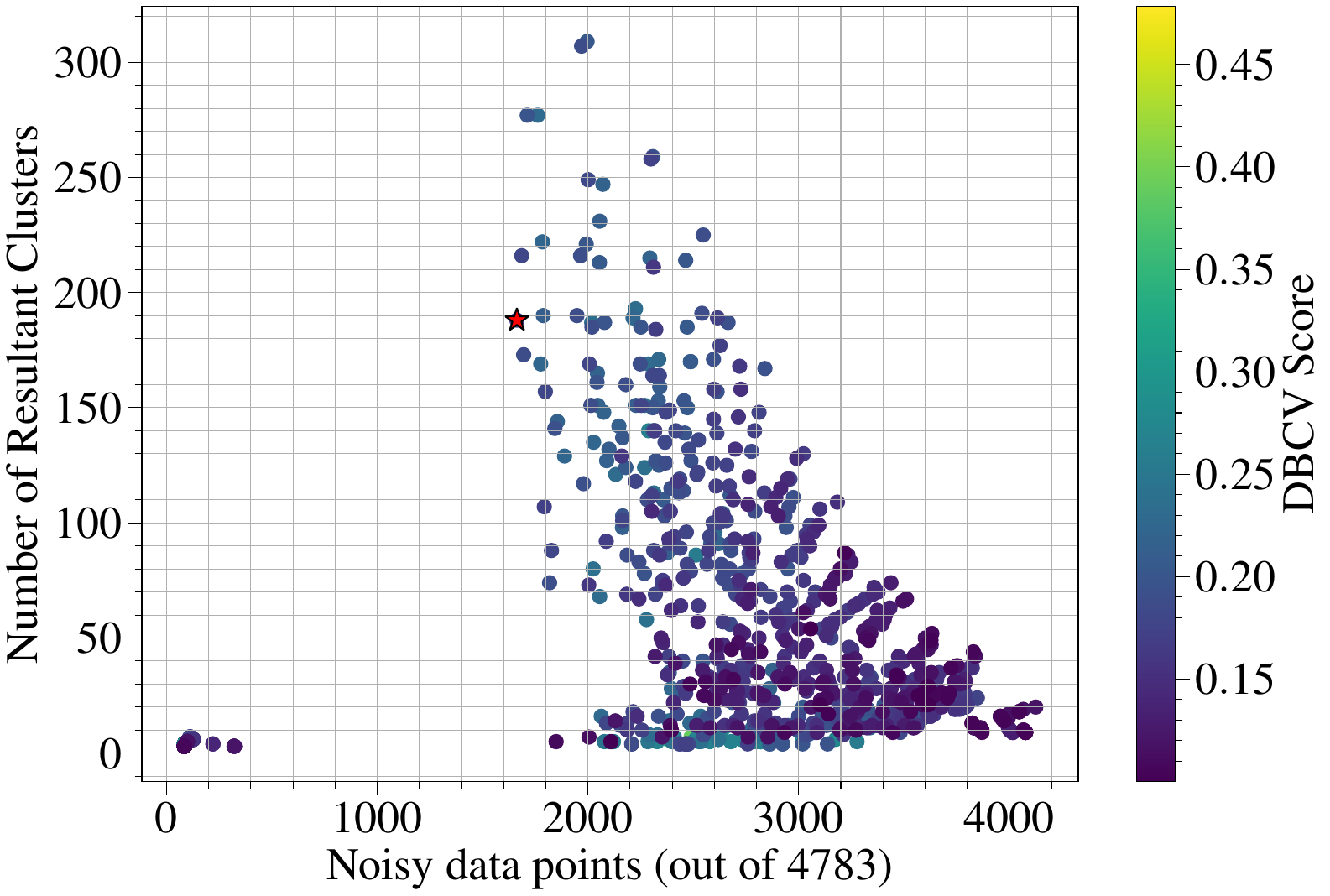}
    \caption{Scatter plot showing number of clusters and noisy data points obtained for different hyperparameter combinations (with DBCV score $>=$ 0.1) from HDBSCAN for clustering of violating adult ads. Red star represents the selected optimal combination.}
    \vspace{-5mm}
    \label{fig:clustering-params-sexual}
\end{figure}

\noindent\textbf{Content Analysis.} To characterize the content of the adult ads, we follow a textual clustering approach employed by Hoseini et al.~\citep{hoseini2023globalization}.
We use a pre-trained multilingual BERT model (distiluse-base-multilingual-cased-v2)~\citep{reimers-2019-sentence-bert} to generate 512-dimensional embedding for each violating ad. 
We selected this specific model primarily because it supports 50+ languages and has been shown to perform well in standard semantic similarity tasks.
Next, embeddings are reduced to 128-dimensional space using UMAP~\citep{mcinnes2018umap} for effective density-based clustering. We use HDBSCAN to perform clustering of these embeddings. We experiment with different combinations of 4 hyperparemeters -- $min\_cluster\_size$, $min\_samples$, $cluster\_selection\_method$, and $metric$ and compute DBCV (Density-based Clustering Validation) score for each combination to evaluate quality of resultant clusters. DBCV score signifies relative density connection between pairs of points and ranges from -1 to 1, where a higher value suggests better clusters. We discard combinations with DBCV $<$ 0.1 as we observe poor clustering results for these cases. Figure~\ref{fig:clustering-params-sexual} depicts the rest of the 1748 combinations. We select the combination ($min\_cluster\_size$=7, $min\_samples$=2, $metric$=euclidean, $cluster\_selection\_method$=eom) that produces the highest number of clusters (188) with least noise (1663 points).
We validate the effectiveness of our 0.1 threshold by selecting two random samples of 100 ad tweets each from the resultant clusters (such that there is at least one ad tweet from each cluster) -- one with DBCV $\geq$ 0.1 and one with $<$ 0.1, respectively.
Next, the primary researcher on the project assign a cluster label to each ad tweet given an ad's textual content and list of cluster labels. Model assigned cluster labels are hidden from the annotator during the annotation to avoid bias. 
Annotation result shows that 60/100 ads with DBCV $\geq$ 0.1 are accurately assigned the correct cluster label as output by BERT, while only 19/100 ads with cosine $<$ 0.1 are correctly assigned, establishing confidence in our threshold.

\begin{figure*}[t]
\centerline{\includegraphics[width=0.9\linewidth]{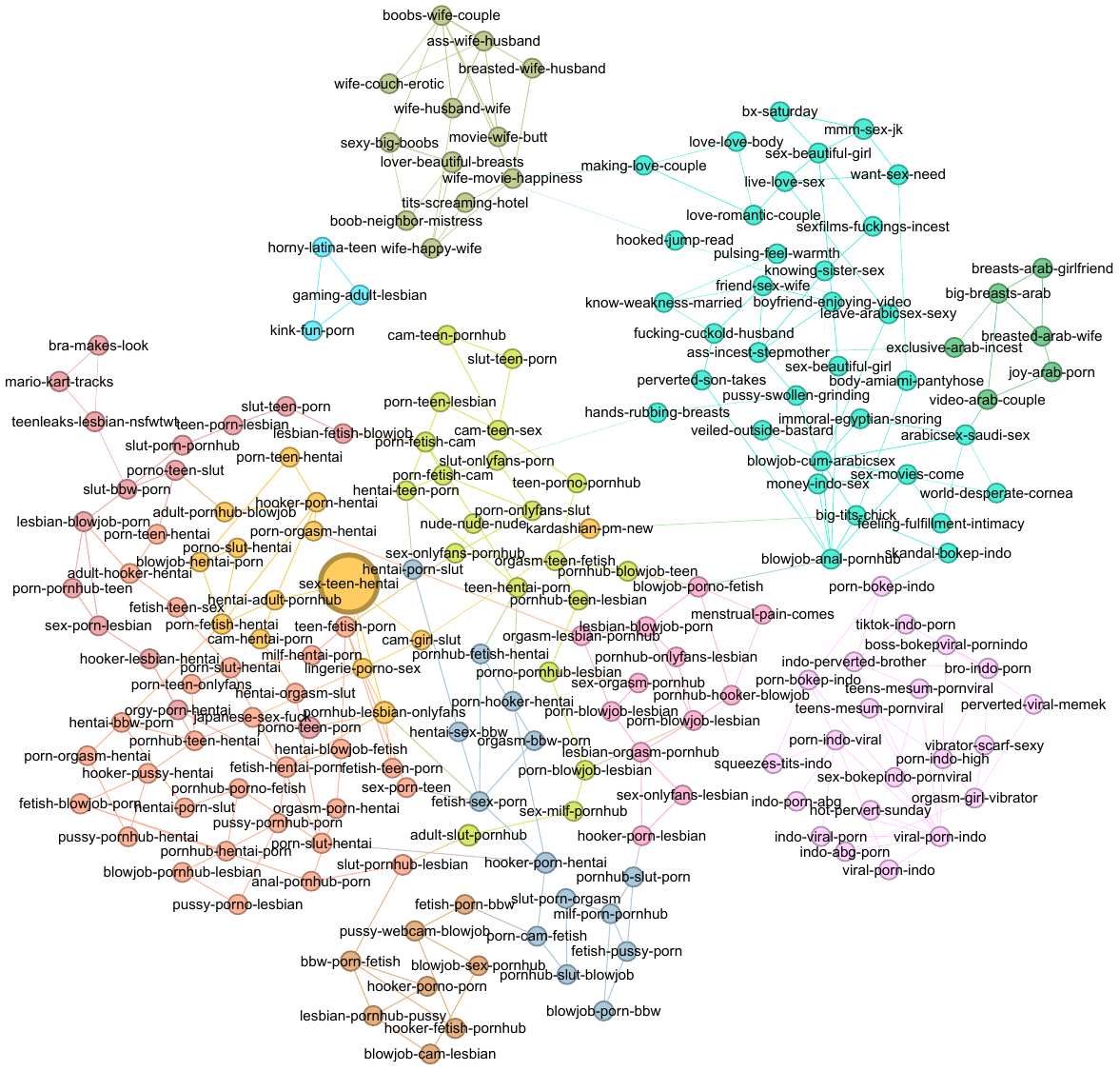}}
\caption{Clustering of sexually explicit ads in violation of Twitter's adult content advertising policy.}
\label{fig:sexual-clustering}
\end{figure*}

Figure~\ref{fig:sexual-clustering} represents the output clusters along with the detected Louvain communities. Qualitative analysis of ads corresponding to different clusters clearly reveals grouping of ads based on distinct adult content-specific categories -- for example, the light purple community in the bottom right represents all the ads that promote \textit{indo-porn} content.
Upon manually evaluating the ads in the violating clusters, we observe that there are 2 main types of content structures followed by the advertisers:
First, the textual content of the ad begins with a period (i.e., ``.'') followed by different space-separated adult content-related words. 
Maddocks et al.~\citep{maddocks2020deepfake} observed this type of bot-based text pattern in normal tweets. 
However, it is surprising to see this pattern in Twitter ads, as it suggests a lack of effective ad moderation. 
The majority of ads with this structure only contain textual content and do not have any embedded links or images. 
3,122 out of 4,873 (i.e., $\sim$64\%) unmoderated adult ads belonged to this category. 
Upon analyzing the usernames of the advertiser accounts in this sub-type, they all followed the same naming pattern -- some hypothetical first name followed by last name without any spaces and first letter of both first and last names capitalized, for instance \textit{MilliexChanax}. 
Upon mapping different advertisers' \textit{author\_id} to their \textit{username}, we observed a pattern wherein to avoid detection, an advertiser account (i.e., a given \textit{author\_id}) kept changing their \textit{username} in variable amounts of time. We observed as high as 49 \textit{usernames} mapped to a single \textit{author\_id}. 
It is important to note that the characteristics of non-moderated sexual ad content described above are also followed by the moderated adult ads discussed in Section~\ref{sec:removed-results}. However, despite that, Twitter is ineffective in moderating all ads that follow the same content pattern.
The remaining ads contain some textual content describing the scenario displayed in the linked porn video, and the catchy text is aimed to entice the viewer's attention using clickbaity adult language. 
This sub-category of ads involved scenarios where the advertisers used dark patterns and embedded unsafe links in violation of Twitter's quality policy~\citep{TwitterQualityPolicy} for advertising, as explained later in this section. 
A standard embedded video within a tweet {should always be} playable on Twitter itself {when clicked on the play button} as per Twitter's policy. 
However, we observed that some advertisers embed a link in the ad with a preview containing an image with a play button to deceive the viewers into thinking it is a normal playable video so that when they click on the play button, it redirects them to the embedded link. 
Twitter moderates even the videos. 
So, to evade detection, some advertisers were found to embed an adversarial playable porn video which was edited so as to have a static foreground overlaid on the video such that humans could perceive the video contents. 
However, the moderation framework could falsely classify it as negative. 
These tactics are strictly against Twitter's quality policy. However, specific behavioral insights of non-compliance discussed here can help platform X in improving its ad moderation and compliance framework, as ads with these deceptive or evasion techniques were never moderated by Twitter.
Moreover, Twitter allows posting of adult content but limits its spread on the platform. However, such porn-disguised-as-ad can potentially be promoted like a normal ad to reach a much larger audience, increasing the harm to vulnerable users.

\begin{figure}[t]
\centerline{\includegraphics[width=\linewidth]{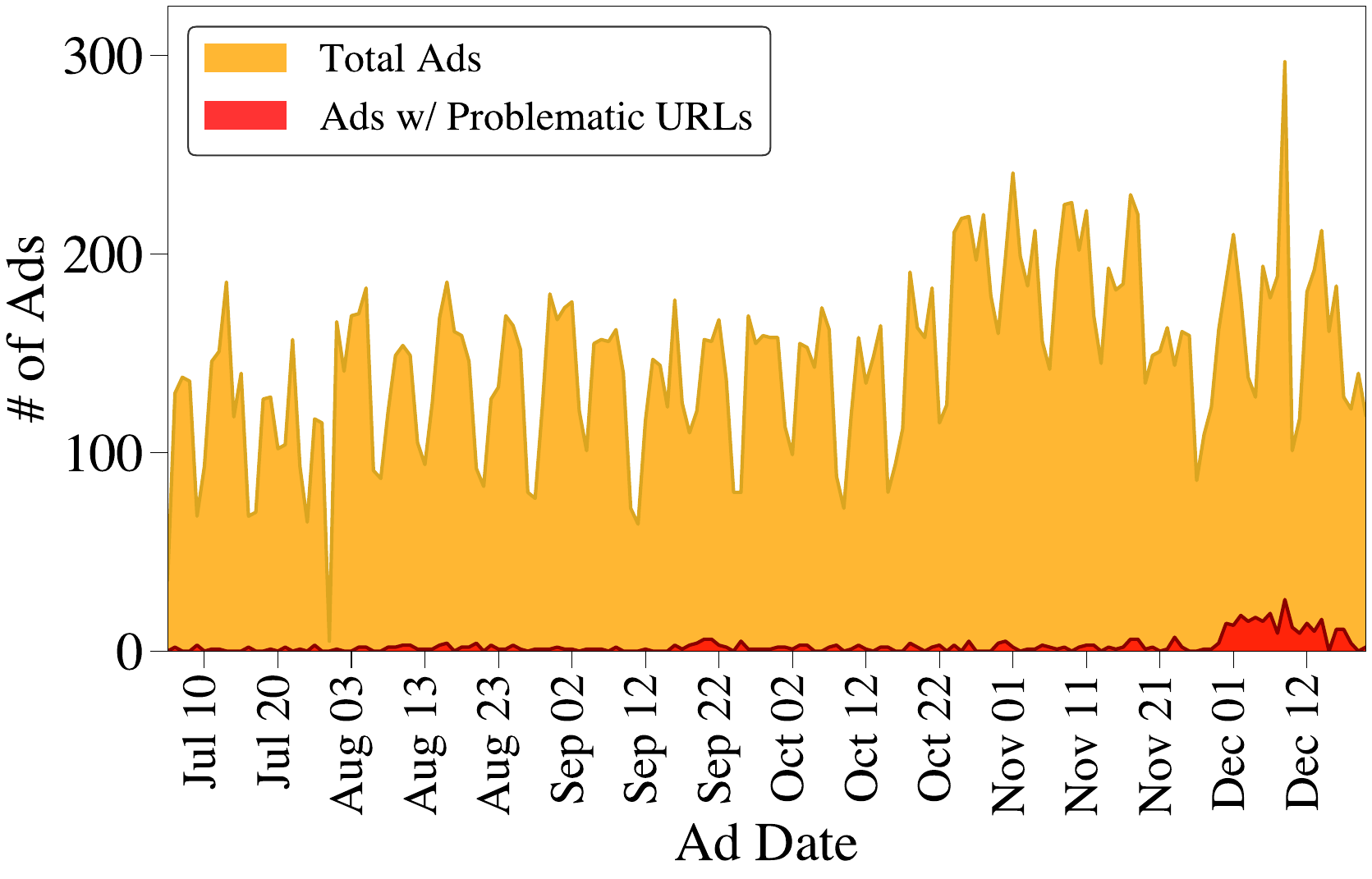}}
\caption{Distribution of the problematic ads with respect to all the ads run on each day on Twitter. Here, problematic refers to classifying either the embedded or the landing URLs of an ad as malicious or suspicious or both by 3 or more VirusTotal services.}
\label{fig:sexual_ads_versus_daily_distribution}
\end{figure}

\noindent \textbf{URL Analysis.} Out of 24,530 non-moderated ads, 20,362 ads had at least one URL embedded within the tweet. 
We observed a common usage of URL shortening services to shorten the original URL before embedding it. 
Besides the benign functionality of reducing the length of URLs, advertisers could also use it to hide malicious URLs in their ads. 
As a result, we analyze the security of embedded as well as the final landing URLs resulting from clicking on an embedded link. 
Using the VirusTotal-based classification of each URL, we calculate a score as follows: 
\begin{equation}
    \label{eq:problematic_tweet_equation}
    \centering
    Score = max(mal_{e}+sus_{e}, mal_{l}+sus_{l})
\end{equation}

Here, $mal_{e}$ and $mal_{l}$ are the counts of VirusTotal services that respectively classified the embedded and landing URL of the current tweet as malicious. 
Similarly, $sus_{e}$ and $sus_{l}$ are the respective counts of VirusTotal services that respectively classified the embedded and landing URL of the current tweet as suspicious. 
Then, we treat a tweet as problematic if the above score is greater or equal to 3 for at least one URL embedded in it. 
We experimented with different values for this threshold -- with a threshold value of 2, many benign links were also getting classified as problematic; with higher thresholds above 3, there was up to 11\% decrease in problematic tweets with an increase in the threshold from 3 to 7. 
As a result, we chose the score threshold as 3 to avoid false negatives. 
Figure~\ref{fig:sexual_ads_versus_daily_distribution} depicts the number of ad tweets with problematic URLs with respect to the daily total ads after the rehydration\endnote{{Moderated ads are no longer accessible, we could not perform this analysis on removed ads.}}. 
We observe an increase in ads with problematic URLs from December 2022 onwards.
This is likely due to the fact that amnesty was granted by Elon Musk to all the past suspended Twitter accounts -- such as that of the former US president Donald Trump, for instance -- by restoring them.
Another benign yet worrisome reason could be that Twitter is unable to detect ads malicious URLs in two weeks, and moderation of such ads requires longer. This explanation can be deduced based on the temporal distribution of URLs in Figure~\ref{fig:sexual_ads_versus_daily_distribution}.

\begin{figure}[t]
\centerline{\includegraphics[width=\linewidth]{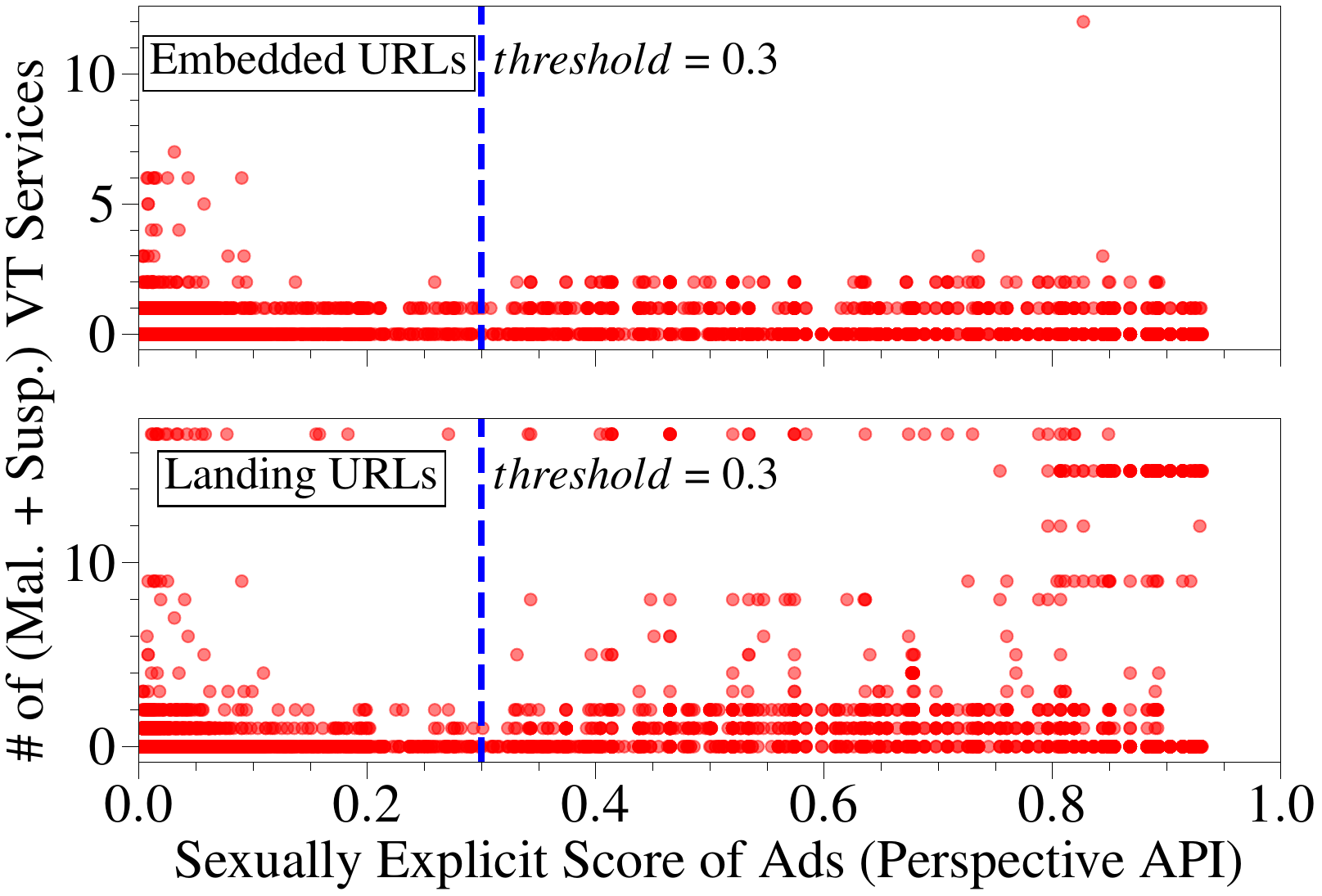}}
\caption{Scatter diagram of (malicious + suspicious) counts of VirusTotal (VT) services for embedded URLs and landing URLs partitioned with respect to the threshold of \textit{sexually\_explicit} score (i.e., 0.3) of the ad tweet containing those URLs.}
\label{fig:scatter_sexscore}
\end{figure}

It is important to understand how different URL usage is for violating adult ads compared to other ads. 
As a result, we plot a scatter chart between the sum of (malicious + suspicious) VirusTotal services and \textit{sexually\_explicit} scores of embedded URLs and Landing URLs in Figure~\ref{fig:scatter_sexscore}. 
Let's first look at the non-adult ads (i.e., \textit{sexual\_explicit} $<$ 0.3) -- for both embedded and landing URLs, we obtain 103 ads where its URL is problematic. 
For adult ads (i.e., \textit{sexual\_explicit} $>$= 0.3), in the case of embedded URLs, only 3 ads contain a problematic embedded link, while on the other hand, 345 ads contain benign embedded URLs, which lead to unsafe landing pages. 

Upon closely analyzing different kinds of landing URLs, we observed a variety of potentially harmful scenarios. 
First, the landing URL sometimes results after a number of redirects. 
Moreover, Twitter only expands the embedded URLs, which are converted by Twitter's link service, and warns its viewers about any harmful page by checking it against a list of potentially dangerous sites it maintains before proceeding~\citep{twitterlinks}. 
However, it does not check the landing URLs currently. 
Landing URL pages comprise -- normal porn sites, fake reCAPTCHA and consent options, system-infected warning screens, anti-virus download prompts, online game sites (including interactive adult games), sex-baity sites, YouTube channel promotions, Forex market trading publisher websites, betting websites (like bet365 offers and registration), etc.

\noindent \textbf{Takeaways.}
We summarize the takeaways from our audit of Twitter's adult content advertising policies with respect to non-moderated ad content:
\begin{itemize}
    \item We observe $\sim$37\% of all sexually explicit ads in our dataset to bypass Twitter's moderation. They constitute 20\% of all the rehydrated ads and are in violation of Twitter's \textit{adult sexual content advertising policy}.
    \item Majority ($\sim$64\%) of non-moderated adult ads follow a simple ad template comprising of a `period' followed by a set of adult words. Advertisers running such ads are suspected to have a higher tweets-per-minute ratio than a normal account, suggesting automation. Despite similar characteristics, Twitter is able to moderate such ads only partially.
    \item Adult ads with videos and embedded links are observed either using dark patterns to engage users or using problematic links. A non-negligible fraction ($\sim$2.75\%) of daily ads on an average are detected to contain problematic URLs (as per VirusTotal) -- most of which embed benign shortened URLs that actually lead to malicious landing pages. This pattern is more common in adult ads than in other ads.
\end{itemize}

%% file: conclusion.tex
\section{Discussion}

In this work, we performed a large-scale audit of Twitter compliance and enforcement of Twitter's adult content advertising policy. 
We selected Twitter because it is understudied in terms of compliance and enforcement, and unlike most social media platforms that strictly prohibit both posting and promoting adult content, Twitter permits posting but prohibits promotion of such material.
We collected a total of 35K advertisements on Twitter and analyzed them to shed light on how Twitter moderates ads containing adult content and the degree of compliance with the platform's adult content advertising policy.
Among other things, our study finds that a large percentage of advertisements on Twitter are sexual in nature ($\sim$38\%) and that 37\% of them are likely not effectively moderated by Twitter. 
Also, we find moderation discrepancies and significant differences in content across various languages; Twitter is more effective in moderating adult advertisements in English when compared to adult ads in Japanese.
At the same time, we find inconsistencies in Twitter's moderation practices; adult ads that follow a simple template (period followed by a list of sexual words) are not consistently moderated and removed by Twitter.
Overall, our work and findings have several important implications for stakeholders interested in online advertising and content moderation. 
Below, we discuss the implications of our work related to data access, content moderation of online advertising, harms that may arise from exposure to advertisements that have a sexual nature, and platform's moderation workflow and affordances.

\subsection{Data Access}

Our work relied on access to the Twitter API -- we leveraged the source attribute that was made available via the API to identify ads and the application used to share or promote tweets.
Unfortunately, during our data collection, our free academic access to the Twitter API and the availability of the source attribute were cut off, mainly due to changes made on the Twitter platform after Elon Musk’s acquisition~\citep{TwitterV2APIMissingSource}. 
These changes to data access have detrimental effects on scientific research efforts that aim to understand and mitigate harmful online phenomena and harm transparency efforts.

Twitter introduced access tiers~\citep{XDeveloperPortal} on February 9th, 2023, under which with a basic tier, researchers get access to only 15K tweets per month at \$200. 
Before that, anyone could access up to two million tweets from the past seven days for free every month. 
Moreover, research institutions could obtain unlimited amounts from the entire archive for free. 
This magnitude of data allowed researchers to rigorously study online communities and analyze platform harms such as bot activities, spread of misinformation, hate speech, political propaganda, etc. at a macroscopic as well as fine-grained level. 
But that is no longer possible. 
As of August 14, 2024, even Meta decommissioned its public insights tool -- CrowdTangle~\citep{MetaCrowdTangle}. 
Without access to data, there is little scrutiny over the platform's practices. 
There is need for a regulatory change that requires platforms to provide data access to researchers.

We argue that, as a research community, we need to work closely with policymakers to ensure that the platforms are accountable for harms arising from their services and must provide the necessary means for researchers to audit their services. 
Indeed, recently, Digital Services Act~\citep{dsa} introduced by the EU Commission went into effect, which calls for such audits on social media platforms to understand if and how their advertising systems are manipulated in the EU. 
We believe this is an essential step in the right direction. 
At the same time, we believe there is still a long way to go about finding ways to incentivize online platforms to be more transparent and to give data access to researchers.

\subsection{Content Moderation}

With his acquisition of Twitter, Elon Musk vocalized his stance against moderating content that is not strictly illegal~\citep{GuardianMuskFreeSpeech}. 
This view empowers free speech but also opens door to hate speech and other types of malicious activities that may cause more harm than good to the end users.

Our empirical analysis demonstrates the blind spots, inconsistencies, and ineffectiveness of Twitter's moderation practices to prevent the publication of harmful advertisements, in this case, ads promoting adult content. 
Our findings are in line with previous characterizations of Twitter's moderation practices on non-ad content and their inability to effectively and consistently moderate and prevent the sharing of harmful content~\citep{zannettou2021won,paudel2023lambretta}.

Post acquisition, on November 4th, 2022, Elon Musk fired 50\% of Twitter employees~\citep{ReutersTwitterLayoffs} -- this constituted 15\% of its moderation team~\citep{CBSMuskLayoffs}. 
This was followed by two additional lay-offs on November 15th and 19th, which worsened the platform's content moderation as per its policies at the time. 
We observe this in our data based on the increase in the net volume of tweets as well as the volume of removed content around this time in Figure~\ref{fig:tweet_distribution}. 
Figure~\ref{fig:sexual_ads_daily_distribution} also depicts the increase in volume of adult ads during this time. 
The situation further aggravated when amnesty was granted on December 1st, 2022 to all the users who were either banned or suspended under Twitter's regime~\citep{GuardianMuskAmnesty}. 
Twitter also revised its adult content advertising policy to allow promotion of certain products that were previously disallowed\footnote{These changes were observed from Oct 2024 as per the Wayback archive.}. This included dating and/or marriage brokerage services, sexual Merchandise (for example sex toys, sexualized clothing), condoms, lubricants and other non-prescription contraceptives~\citep{RevisedBusinessAdsPolicy}.
This shows that already poor compliance and moderation got worse with laxed enforcement of policies and lack appropriate moderation staff.

Overall, our findings emphasize the need for effective and consistent content moderation, especially in the context of online advertising. 
Even for online platforms, this is important because ineffective and inconsistent moderation of advertisement content can have negative consequences. 
For example, advertisers can choose not to advertise on the platform due to brand-safety concerns~\citep{TimeTwitterHateContent} or users loosing trust in the platform's ability to diminish the propagation of harmful content and migrate to other platforms~\citep{ForbesDeleteX}. 
Due to this millions of users have migrated to newly emerged platforms -- Mastodon and BlueSky -- for better moderation~\citep{NewsweekTwitterRivals}

Twitter's adult content advertising policy prohibits promotion of adult content globally. 
Specifically, policy clearly prevents promoting ``\textit{sexually suggestive text, images, videos, audio, gestures, poses and movements etc.}'' 
Our results show that not only videos with adult content go undetected from advertising but also that the platform is unable to moderate simple ad tweet text with a bunch of sexual words that is promoted. 
This is different from all other social media platforms that generally disallows both sharing as well as promoting adult content on their platform. 
One of the other important distinction is that Twitter also provides a capability of advertising any post that was originally created as a normal tweet for a brief period of time. 
This is crucial due to the unique adult content advertising policy under which a malicious advertiser can create a normal adult content tweet and then later on promote it. 
Our approach cannot detect such cases as the data obtained using Twitter's API doesn't provide any information that suggests if the tweet was ever promoted in its lifetime or not.

Additionally, our results highlight significant differences in the moderation of advertisements across languages, which demonstrates the need to improve content moderation in lower resource languages (when compared to English). 
That is, content shared in non-English languages is more likely to bypass content moderation, which emphasizes that online harm can be more prevalent on content that is not shared in English.
These findings have important implications and emphasize the pressing need to improve content moderation and moderators' workforce that is able to effectively deal with content shared in languages other than English. 
We discuss more about the platform's moderation workflows in the Section~\ref{sec:workflows}.

\subsection{Potential Harms}
Trust and safety are crucial to evaluate and then appropriately mitigate harms arising from services offered by social media platforms.
Our study shows that there is a non-negligible amount of adult ads on Twitter. 
This also includes porn promoted as an ad by bad actors.
Its presence on platforms could potentially harm curious youths' image and perception regarding sexual relations and sexuality~\citep{paasonen2024nsfw}.
According to a study, one in five youth between the ages of 9 and 17 would view unwanted sexual material online~\citep{OneInFiveYouth}.
Also, nearly 25\% of youth presented with adult content experience intense fear or anxiety~\citep{madigan2018prevalence}.
Such inappropriate content reduces trust in digital platforms. 
When platforms like Twitter fail to adequately moderate adult ads, it sends conflicting signals to users regarding the platform's commitment to user safety. 
This can also disproportionately harm marginalized groups, such as women and LGBTQ+ youth, who are more likely to encounter stigmatizing or exploitative depictions of sexuality online~\citep{powell2020image}.
A study found that increasing the intensity of sexual humor in advertisement improves its effectiveness amongst men but leads to significantly more negative attitudes toward the advertisement and brand amongst women~\citep{freeman2023would}.
Stanford researchers showed an increase in CSAM content on the platform in 2023~\citep{StanfordIllicitContent} where illicit adult content was distributed by a network of accounts operated by minors.
Moreover, Participants of a user study focused at understanding user perspectives on content of an ad, have expressed disgust against the presence of sexual content in ads~\citep{zeng2021makes}.

Besides increasing risk of its exposure to vulnerable platform users, advertising adult content could also result in the potential exploitation and trafficking of vulnerable individuals through poorly moderated ad spaces. 
Studies have highlighted that online platforms, including social media, have been exploited by traffickers to advertise exploitative services, often under the guise of adult content or escort ads~\citep{latonero2011human}. 
When platforms like Twitter fail to regulate such ads, they risk becoming unwitting enablers of trafficking networks, where victims of exploitation are marketed and sold to large audiences with minimal accountability~\citep{williams2019trafficking}. 
Furthermore, the global reach of social media platforms exacerbate the scale of this harm, making it harder for authorities to track and intervene. 
Overall, there is a pressing need to improve platform affordances and moderation workflows so that these potential harms are effectively mitigated from online platforms like Twitter.
\subsection{Moderation Workflows}
\label{sec:workflows}

To better understand ineffective moderation, it is important to look into Twitter's moderation workflow~\citep{ForbesAIModeration}. 
Twitter adopts a two-fold approach in its ad moderation constituting of automated systems and human moderators. 
In the first phase, Twitter's automated systems perform a large-scale review of platform-wide posted content, including ads using AI-based approaches designed to detect violations such as hate speech, violence, inappropriate content, etc. 
If these automated systems flag an ad to be problematic, appropriate human moderators review the ad content to reach a decision about whether to moderate the content or not. 
A label is assigned to an ad to provide context and additional information to the users. \citep{XRulesEnforcement}. 
Then, on a case-by-case basis, it maybe either restricted or the account maybe suspended/terminated.

Our research shows that the adult ads are disparately moderated across different languages, suggesting that sufficient number of language-specific human moderators are not available. 
This further worsened when the platform fired a portion of its moderation team~\citep{WiredModerationSystem}, leaving with no option but to heavily rely on automated AI systems, lowering platform's  moderation accuracy. 
Recruiting expert human moderators in proportion to the volume of tweets in different languages can help.

We argue that if we can determine that a significant portion of the non-compliant content goes unmoderated using Perspective API, then Twitter has a lot more resources and is more capable (than us) to determine whether problematic content is being shared on the platform. 
A recent study showed that state-of-the-art generative AI models perform better harmful content labeling than crowd workers~\citep{jo2024harmful}. 
Given that Twitter recently updated its platform policies to allow the user posted content to be accessed and utilized to train its generative AI model -- Grok \citep{XAboutGrok}, it can be used to detect and moderate problematic content such as adult content on the platform. 
We finally conclude with recommendations to improve these moderation workflows. 
Existing moderation workflow is able to moderate some ads with a given pattern but not all. 
Lookalike-based moderation models can help moderate these basic non-compliant ads at the very least in a timely manner. 
Additionally, most of the false negatives (although small in number) comprises of an ad with textual content that can be interpreted differently -- sometimes benign appearing text were observed to be associated with additional modality of content that is in violation of adult content policy. 
To effectively handle such cases, context-aware multi-modal models could prove to be useful to incorporate broader context, such as cultural norms, regional variations, etc. 
Local moderators can help moderate flagged content more effectively than a generic moderator. 
Lastly, integrating ad policy checks directly when an advertiser creates an ad can prove to be a low-cost and effective way to early-detect violating ads preventing them from running as part of the campaign. 
This also ensures that advertisers understand and comply with platform advertising policies strictly.